\documentclass[12pt,preprint,flushrt]{aastex}
\usepackage{txfonts}
\usepackage{graphicx}

\begin{document}

\title{UNICS - An Unified Instrument Control System for Small/Medium Sized Astronomical Observatories}

\author{Mudit K. Srivastava$^1$, A. N. Ramaprakash$^2$, Mahesh P. Burse, Pravin A. Chordia,  Kalpesh S. Chillal, Vilas B. Mestry, Hillol K. Das and Abhay A. Kohok}
\affil{Inter-University Centre for Astronomy and Astrophysics (IUCAA), Pune, India}
\email{
${}^1$mudit@iucaa.ernet.in\\
${}^2$anr@iucaa.ernet.in
}

\begin{abstract}
Although the astronomy community is witnessing an era of large telescopes, smaller and medium sized telescopes still maintain their utility being larger in numbers. In order to obtain better scientific outputs it is necessary to incorporate modern and advanced technologies to the back-end instruments and to their interfaces with the telescopes through various control processes. However often tight financial constraints on the smaller and medium size observatories limit the scope and utility of these systems. Most of the time for every new development on the telescope the back-end control systems are required to be built from scratch leading to high costs and efforts. Therefore a simple, low cost control system for small and medium size observatory needs to be developed to minimize the cost and efforts while going for the expansion of the observatory. Here we report on the development of a modern, multipurpose instrument control system UNICS (Unified Instrument Control System) to integrate  the controls of various instruments and devices mounted on the telescope. UNICS consists of an embedded hardware unit called Common Control Unit (CCU) and Linux based data acquisition and User Interface. The Hardware of the CCU is built around the Atmel make ATmega 128 micro-controller and is designed with a back-plane, Master Slave architecture. The Graphical User Interface (GUI) has been developed based on QT and the back end application software is based on C/C++. UNICS provides feed­back mechanisms which give the operator a good visibility and a quick-look display of the status and modes of instruments. UNICS is being used for regular science observations since March 2008 on 2m, f/10 IUCAA Telescope located at Girawali, Pune India.
\end{abstract}


\section{Introduction}
The emergence of large optical telescopes along with state of the art back-end instruments have greatly benefited the astronomical community through their unique capabilities to see deeper in the space. However the relevance of smaller and medium size optical telescopes continue to grow considering the smaller number of the large new facilities and the ease of availability of telescope time on such telescopes. In order to minimize the time overheads during the observations in the setup of various instruments and units (like filters, calibration lamps etc.) and in the data acquisition, it is essential to include modern technologies in the conception of new instruments and in control and data acquisition systems. However unlike large optical telescopes where focal plane instrumentation as well as their control systems have reached fairly high level of complexity and are more expansive in nature, the control systems for smaller and medium size observatory are required to be simple in nature.
\par
Here we shall describe the design and development of such an control system named {\bf `UNICS - UNified Instrument Control System'} which has been developed for IUCAA Girawali Observatory (IGO), that houses a modern 2m, f/10 Telescope at Girawali near Pune India \citep{SNT1998, gupta2002}, but fulfills the requirements of any small and medium size (less that 4 meter diameter Telescope) Observatory. A modern optical observatory consists of various instruments, devices and hardware systems and demands sophisticated control and monitoring of these systems during observations. Therefore it is extremely desirable to integrate the controls of these systems within a single framework to reduce the complexity, cost and efforts to maintain the system and to provide a single interface to the user. UNICS is developed keeping the above objectives in picture. The whole structure of UNICS can be thought of consisting two separate environments: (1.) Linux based Graphical User's Interface (GUI) and Application Software (2.) A micro-controller based Embedded Control System, named Common Control Unit (CCU) and its embedded software. These two parts of UNICS communicate to each other through a serial link or through an Ethernet based communication link for remote operation. 
\par
This paper describes the design and development of UNICS to fulfill the above discussed requirements and various novelties involved at various level, in the context of automated astronomical observations at small or medium size optical observatory. In particular the Embedded Control Unit (CCU), Application Software and the graphical user interface (GUI) are discussed here. Section 2 provides the design and implementation plan of UNICS. The basic requirements and ways to handle them are discussed here. Section 3 is the description of the CCU. Its architecture and various software issues (inter processor communication, protocol etc.) are described here. Section 4 consists of the description of the GUI and Application software. Section 5 summarizes the paper and performance of UNICS is also presented in this section.

\section{UNICS Requirements, Design \& Implementation}

\subsection{UNICS : General Requirements and Considerations}

Like any other control system, the prime role of UNICS is to provide access of various instruments and devices mounted on the telescope to the end user in the control room; so that they can be monitored and controlled accordingly during the observations. Further, considering possible future expansion of the observatory both in terms of new focal plane instrumentation as well as the replacements of old devices, it was extremely desirable to have a control system that should be flexible and scalable in nature thus allowing the reconfiguration of existing facility and integration of newer devices or instruments with minimum efforts. As the aim was to develop a control system for medium size astronomical observatories, additional constrain from financial side were also there in picture. Therefore the option of generalized control frameworks from various commercial solution providors (that could be configured with relatively ease) was not feasible and hence the emphasis was given to develop an in-house framework that could be replicated and expanded with little extra effort and cost. Also during the conception of UNICS the focus was to use inexpensive platforms (both in software and hardware), as well as to adopt the accepted industry standards, so that it could be easily maintained in the future; and hence UNICS software was developed by using open license systems. Further to reduce the complexity of the system the number of different hardware and software components were minimized by using only single architecture while designing the hardware and software of UNICS. This approach will be discussed in the later sections. Lastly, another very important objective was to develop a user friendly environment via a GUI which would provide an efficient and consistent control of the facility.

\subsection{UNICS : Scientific and Technical Objectives}

The factors mentioned above determined the wide scope of UNICS. However the basic technical requirements of this Unified Control System originated from the idea to integrate high level controls of various instruments, detectors and data acquisition in a single system. For example, the present instruments on IUCAA Telescope include IUCAA Faint Object Spectrometer and Camera (IFOSC) with a Polarimetric mode \citep{gupta2002, ram2002}, Near Infrared PICNIC Imager (NIPI, an infrared imaging system) \citep{ram2003},and a Princeton Instrument CCD Imaging System (PI CCD). Also there are some support systems e.g. a Calibration Unit (consisting of various lamps which are used for spectral calibration purpose) and a power supply unit to control the basic power requirement of these systems. Many of these systems have movable parts (that are controlled through motor drives), switching circuitry, and feedback arrangements which are being used in various ways for automated astronomical observations. These processes were required to be integrated in such a way so that the end user can make the observations through a Graphical User Interface (GUI) without knowing the system in depth. Also in order to utilize the full potential of the telescope a variety of other instruments are being planned spanning the optical and near-Infrared regime of the spectrum. Hence the issue of incorporating the controls of any other external instruments within the existing control system and GUI has been given careful thought. Further taking a step towards fully automated observations the option of allowing user defined macros was another major objective of UNICS. Additionally, the requirement that UNICS need to be seamlessly integrated with the telescope control system as well as auxiliary control system (like environment monitoring), has also been considered. Therefore UNICS is designed so that it should be able:
\begin{itemize}
\item to control all the movable parts of the instruments and support system at the telescope.
\item to operate all the switches and shutters.
\item to take feedback from various sensors mounted on the telescope and,
\item to talk to control system of any other external instrument/system. 
\end{itemize}
Out of the above requirements the last one is most critical as it provides means to add new instruments in the existing set up. Earlier, motion of various systems on IUCAA telescope were controlled by 6 different stepper motors while the IFOSC has its own controller to drive various motors in it. Therefore the necessary hardware requirements for the UNICS were to have the inbuilt circuitry to drive all the motors, shutter controls drivers for all the shutters on the telescope, some switching relays to control various power supply lines, and ability to gather the status of a number of digital and analog feedbacks. In order to communicate data with external devices and controllers (for example IFOSC's controller) some hardware arrangement were required. All these hardware requirements are combined in form of an embedded Control Unit named {\it Common Control Unit (CCU)}. Serial communication protocols are being used at various levels in UNICS, however remote accessibility is also given to UNICS through Ethernet network link.
\par
At the user end an application software was also required to be developed which could take various instructions from the users through a GUI and pass it to the CCU in a standard way as well as provide an option and standard platform to include the controls of various other instruments and devices. Though the CCD data acquisition system on IUCAA telescope has its own hardware independent of CCU, its software controls were also required to be integrated within the same GUI. Further due to past experiences and to ease in calibrating the system during the engineering tests, debugging, fault diagnosis etc. an engineering GUI interface has been provided which would execute all the basic commands, and where all the required system parameters can be checked and set.

\section{Common Control Unit (CCU)}

\begin{figure}
\centering
\plotone{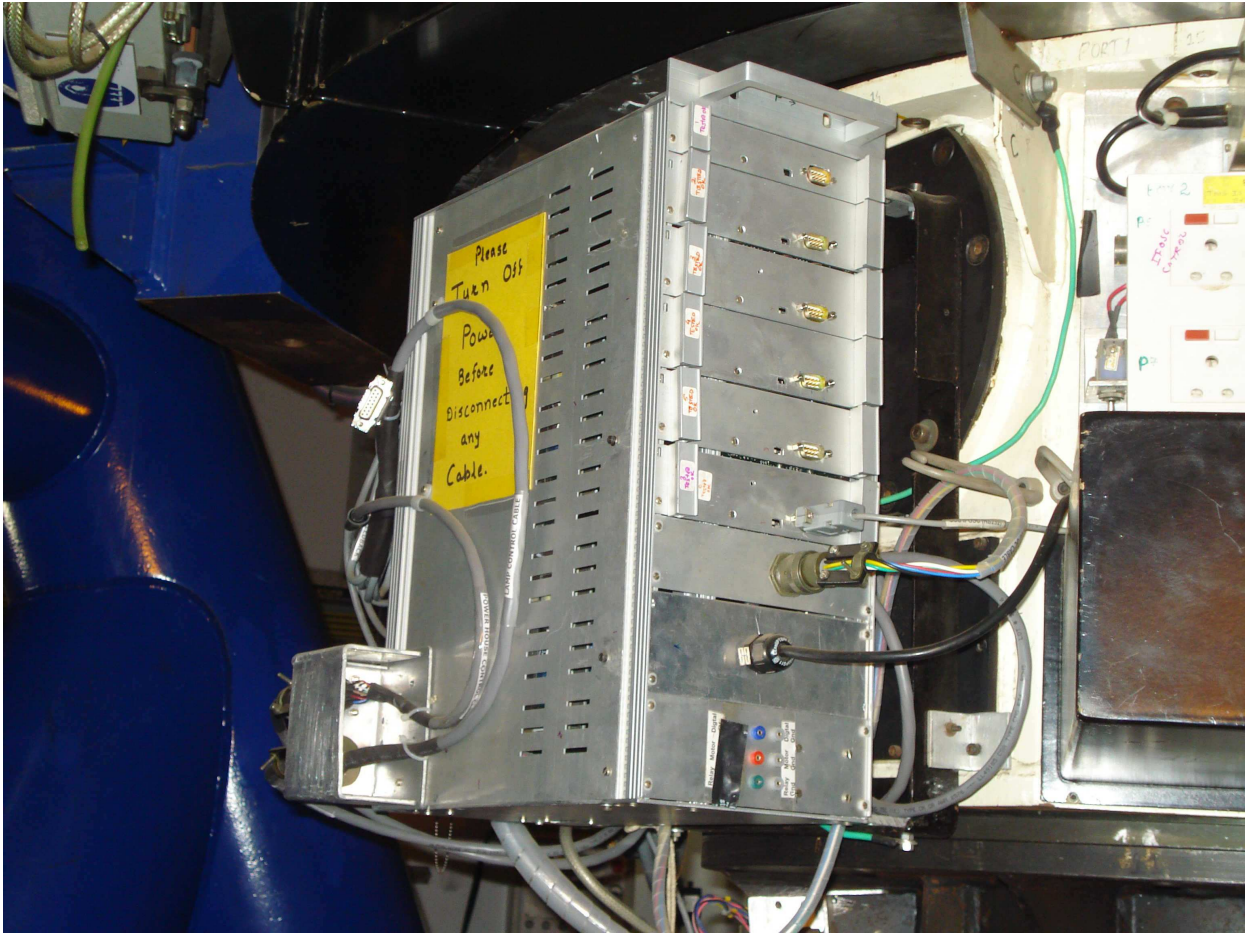}
\caption{The Common Control Unit (CCU) on IUCAA Telescope: CCU is the hardware part of UNICS consisting 6 identical cards on a Backplane card using Master-Slave architecture. It also contains three SMPS power supplies. RS-232 based serial communication as well as Ethernet protocol has been used to communicate with CCU.
}
\label{CCU}
\end{figure}

Common Control Unit (CCU) is the embedded control unit of the UNICS. It is designed and developed to accept various commands and instructions from the user through GUI and execute them. The basic architecture of CCU is built around Atmega 128 micro-controller and the design is based on the Master-Slave Back-Plane architecture. CCU has been developed following the relevant industry standards (while designing the PCB and the enclosure box) for ease in long term maintenance process. The industry standard 3U size is used for the PCB designing while the enclosure box is of the dimension of 4U X 84 T (Figure~\ref{CCU}). Further to reduce maintenance efforts, only one type of card has been developed. This CCU card contains all the required circuitry to fulfill all the functional requirements. The CCU consists of 6 such identical cards and to work in Master or Slave mode any one of the cards can be configured as the Master Card and rest as Slave Cards by changing the settings of few jumpers on the Back-plane Card. As all the CCU cards are identical, any card can be replaced by any other card without making any change in the architecture. The required power supplies for the CCU are also embedded inside the CCU enclosure and three Switch Mode Power supplies (SMPS) are used for this. 

\subsection{CCU : Layout and Hardware Architecture}

\begin{figure}
\centering
\plotone{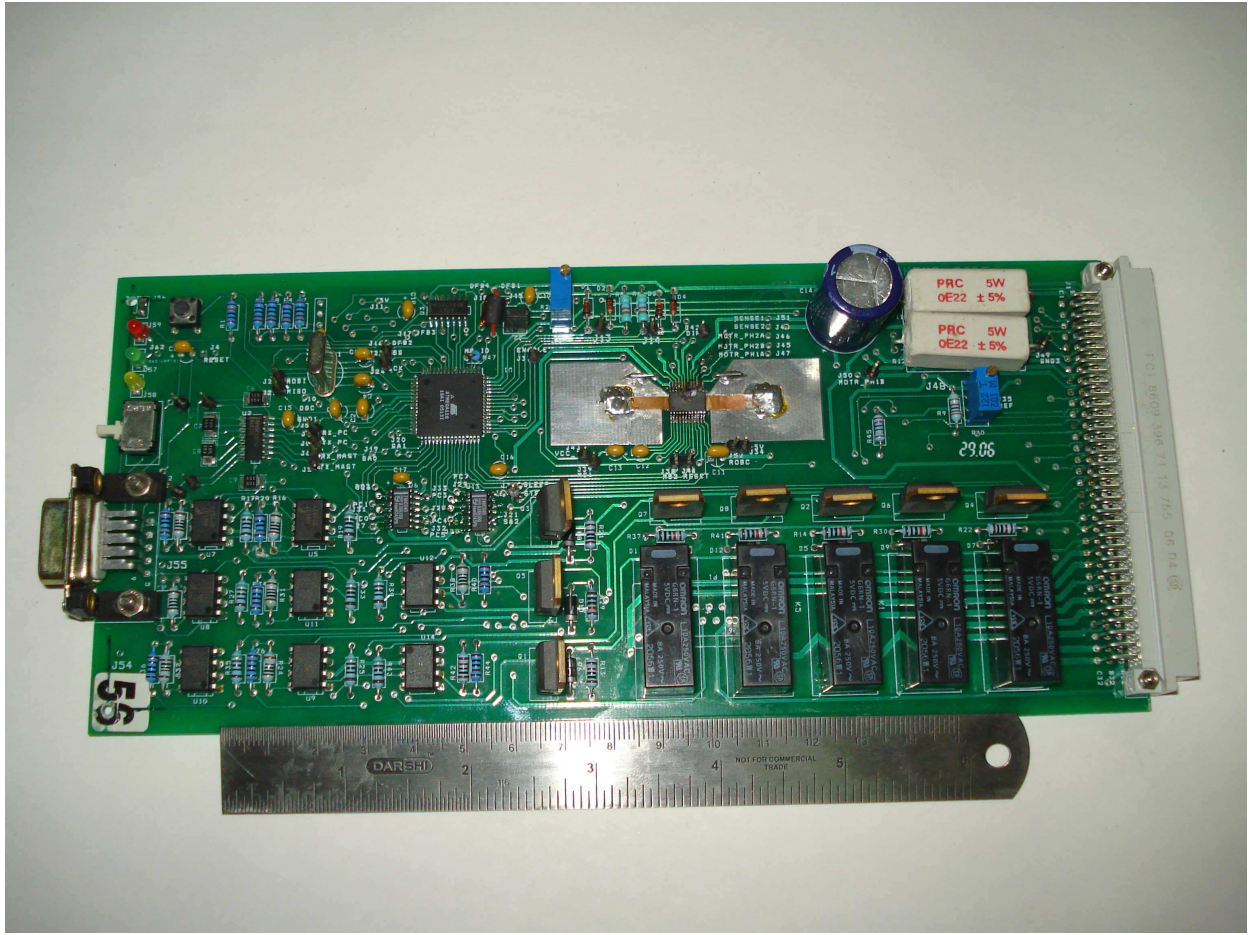}
\caption{PCB card of the Common Control Unit (CCU): This card can be configured into Master card as well as Slave Card depending on the setting of few jumpers on the back-plane card.
}
\label{CCU-Card}
\end{figure}

The basic layout of a CCU PCB card is shown in Figure~\ref{CCU-Card}. At the heart of the CCU is ATmega 128 micro-controller chip from ATMEL Corporation\footnote{http://www.atmel.com/dyn/products/product$\_$card.asp?part$\_$id=2018}. It is an 8 bit micro-controller based on AVR enhanced RISC architecture. ATmega 128 offers various features which make it ideal for our application e.g. inbuilt Serial Peripheral Interface (SPI) for inter-processor communication, Dual Programmable Serial Universal Asynchronous Receiver Transmitter (USART), 8 channel 10 bit ADC etc. Also the availability of 53 programmable I/O lines make it really suitable for the applications where a large number of devices and components are to be controlled or monitored. Further the inbuilt architecture for SPI and USART communication systems have provided a great deal of simplicity in the hardware and software system design. The CCU uses both the USARTs : USART-1 is used for the user's Server to Master Card communication while the USART-2 is used to communicate with any other external system. In-system programming feature is used to load program into the micro-controller. However at the output only one 9 pin connector is used for the in-system programming as well as for USART-2 (to communicate with the external system). Using a DPDT switch any one of these modes can be selected. SPI is being used for Master-Slave communication process. 
\par
Apart from micro-controller another important component in CCU is A3984 stepper motor driver IC from Allegro Micro System\footnote{http://www.allegromicro.com/en/Products/Part$\_$Numbers/3984}. It has a built in translator and is designed to operate bipolar stepper motors in full, half, quarter and sixteenth step mode with an output drive capacity up to 35V and $\pm$ 2 A, thus allowing the micro-stepping mode of stepper motors. Also A3984 has the ability to operate in Slow or Mixed decay mode. A CCU card consists of an ATmega 128 micro-controller, an A3984 stepper motor driver IC, a RS232 driver/receiver communication IC to convert micro-controller signal to RS232 standard, 5 relays from OMRON G6RN series and their driver circuits. Further an external shutter control circuit can also be added to this card using the connections of 2 out of 5 relays. These relays are kept isolated from the micro-controller using opto-isolators and three different SMPS power supplies are used for digital power (5V - 5A), relay power (5V - 5A) and for Stepper Motor driver IC (24 V - 9 A). 
\par
\begin{figure}
\centering
\plotone{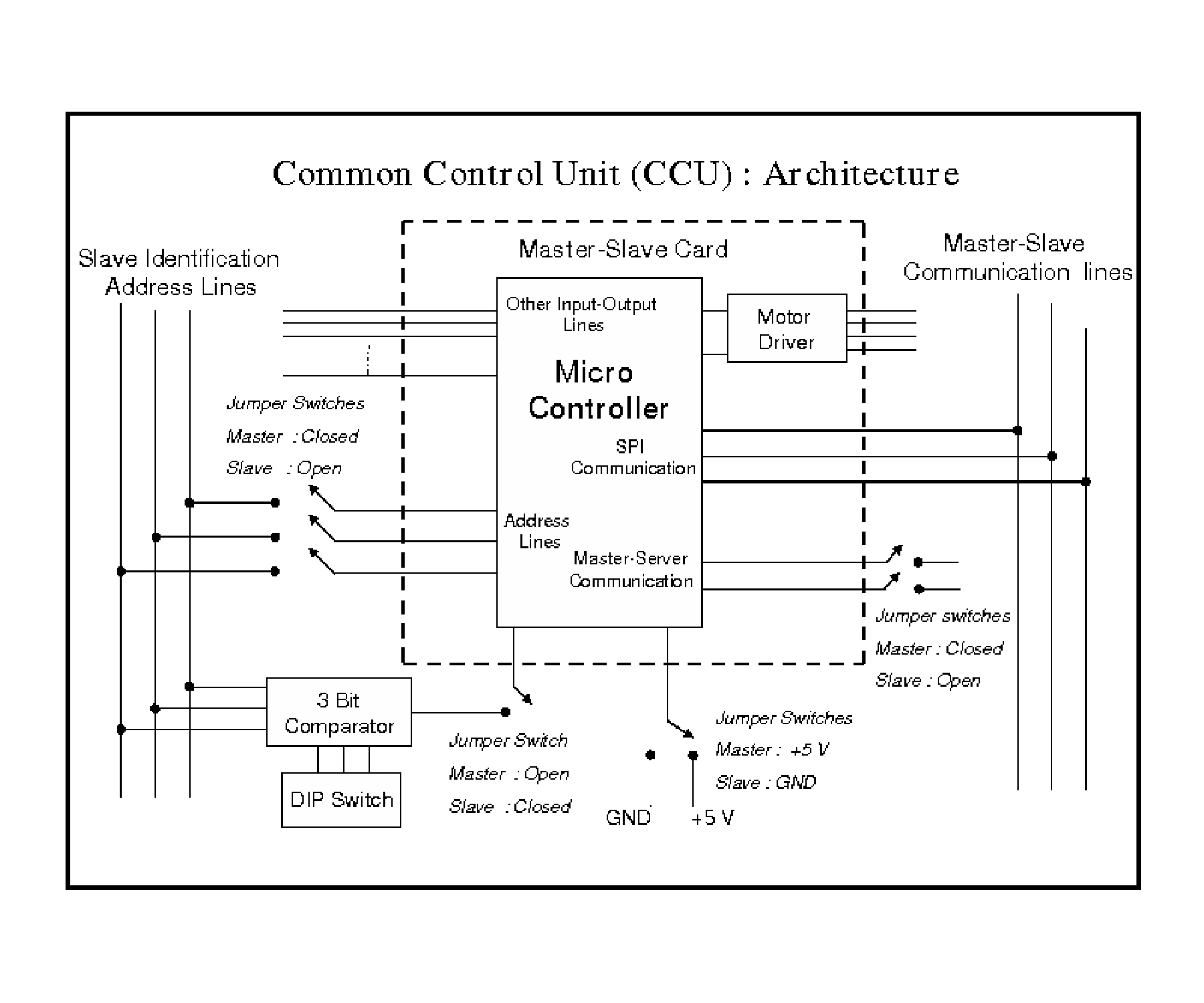}
\caption{The Hardware Architecture of the CCU : The Architecture of a CCU card and its interfaces are shown on the back plane card. There are 6 identical setups are used in the CCU connected by 3 Slave address lines and 3 Master-Slave Communication Lines. The settings of 8 jumper switches on the Back plane Card can be changed to configure a CCU card in Master mode or in Slave mode.
}
\label{CCU-Arch}
\end{figure}

The CCU consists of six cards in Master-Slave configuration and as mentioned earlier this has been implemented in such a way that any card can be configured as Master as well as Slave. The schematic of one card and its external connections on the back-plane card is shown in Figure~\ref{CCU-Arch}. A card could be configured in to Master or Slave mode by changing the logic level of a {\it Master-Slave Identification pin} on the back-plane card and through the execution of an embedded software program. A set of three lines (named as address lines) are used for the purpose of slave addressing and master card would use them to select a particular slave card. Communication between master and slave cards utilizes the inbuilt SPI communication protocol and uses three lines (i.e. Master In Slave Out (MOSI), Master Out Slave In (MISO), and Clock signal) for this purpose. Thus all the six cards of the CCU are connected with each other via these six lines on the back-plane card. Only one of the cards (the master card) can communicate with the server. Moreover same embedded software is loaded into each of the cards. The CCU is finally mounted on the Telescope and fetch power from the mains supply available on the Telescope and is connected to the main server in the control room via fibre-optics cable through RS232 protocol.

\subsection{CCU : Software Architecture and Communication Protocols}

The embedded software architecture of the CCU is responsible for the execution of tasks. Like the hardware architecture, emphasis is given to reduce the complexity of the system. As discussed earlier the hardware has only one basic common structure which is repeated and only by changing the setting of few jumpers and switches on the Back-Plane card every card is configured differently from each other. The same strategy is followed in the development of Embedded Software for CCU. Only one kind of software is developed which is loaded on all the cards of the CCU making every card identical to each other. This helps greatly in the maintenance of the cards as well as in the further development of the software. The Embedded Software architecture of the CCU can be categorized in four parts:
\begin{enumerate}
\item A Command Structure that defines all the basic tasks to be executed and structure of each command at the engineering level.
\item Communication Protocol between the user's Server and CCU. 
\item Communication Protocol between Master and Slave Card. 
\item Algorithms for the execution of various commands.
\end{enumerate}

The Command Structure for the CCU is the basic framework on which the whole UNICS software architecture (both the CCU's Embedded Software and the GUI support) is based. There are in all 28 basic commands in UNICS to control the various parameters of the stepper motors, to monitor and change the status of various switches and sensors and to transfer/receive the data to/from an external instrument or device. Various combinations of these basic commands are constructed at the user's end with appropriate parameters through GUI to execute a task on the telescope (e.g. bringing a particular filter into the path of the beam or turning ON/OFF an spectral lamp for calibration purpose etc.) and hence providing automation to the astronomical observations. A Command is made up of three sections: {\it (a.) Header Part} carries the information of the card's address (System ID) for which the command is meant for, a command code which is specific to each basic command (this will indicate what function is to be performed) and the parameters size that will tell the total number of bytes in the parameter space of the command structure; {\it (b.) Parameter Space} contains the values of the parameters required for the execution of the task; and the last {\it (c.) Trailor part} contains one byte which is sum of all bytes presents in header and parameter sections of the command structure. It is used to check the success of the transmission of the command structure on the other side. As an example the basic structure of the `MOVE' (to rotate the stepper motor by 1000 steps in clockwise direction) command is given below,

\vskip3pt
{
\small\tt

Header : \\
\hspace{5mm}        `system ID' = 3 (1 byte)\\
\hspace{5mm}        `Command Code' = 2 (1 byte)\\
\hspace{5mm}        `Parameter Size'  = 5 (1 byte)\\

Parameters : \\
\hspace{5mm}        `Number of steps' = 1000  (4 bytes)\\
\hspace{5mm}        `Motion Direction' = +1 (Clockwise) (1 byte)\\
       
Trailor :\\
\hspace {5mm}       `Checksum' = 1 byte\\
}

 Appropriate byte packets are generated by the application software using user's input via GUI and sent to the relevant card through the Master's card. The algorithms for Server to Master communication, Master to Slave communication, and for the execution of all the basic commands have been developed. As UNICS is the single node that process the control of all the instruments on the telescope, multithreads processing of the commands, have been incorporated during the development of Embedded Software. This means any given card (whether it is Master or Slave) can execute two jobs at the same time, which is another important feature of the CCU software system. This is in particular helpful in case of Master which should be able to talk to the Server all the time but should also perform some tasks associated with it. Further the possibility to define Macros for the CCU has also been incorporated in to the Application Software as well as in the embedded software of the CCU. Thus various commands with their respective parameters can be set and stored in a sequence in memory of the micro-controller and could be executed at any other time. This would allow users and the engineers to make their own sequence of commands during the observations and routine maintenance of the system.

\section{Application Software and Graphical User Interface}

\begin{figure}
\centering
\plotone{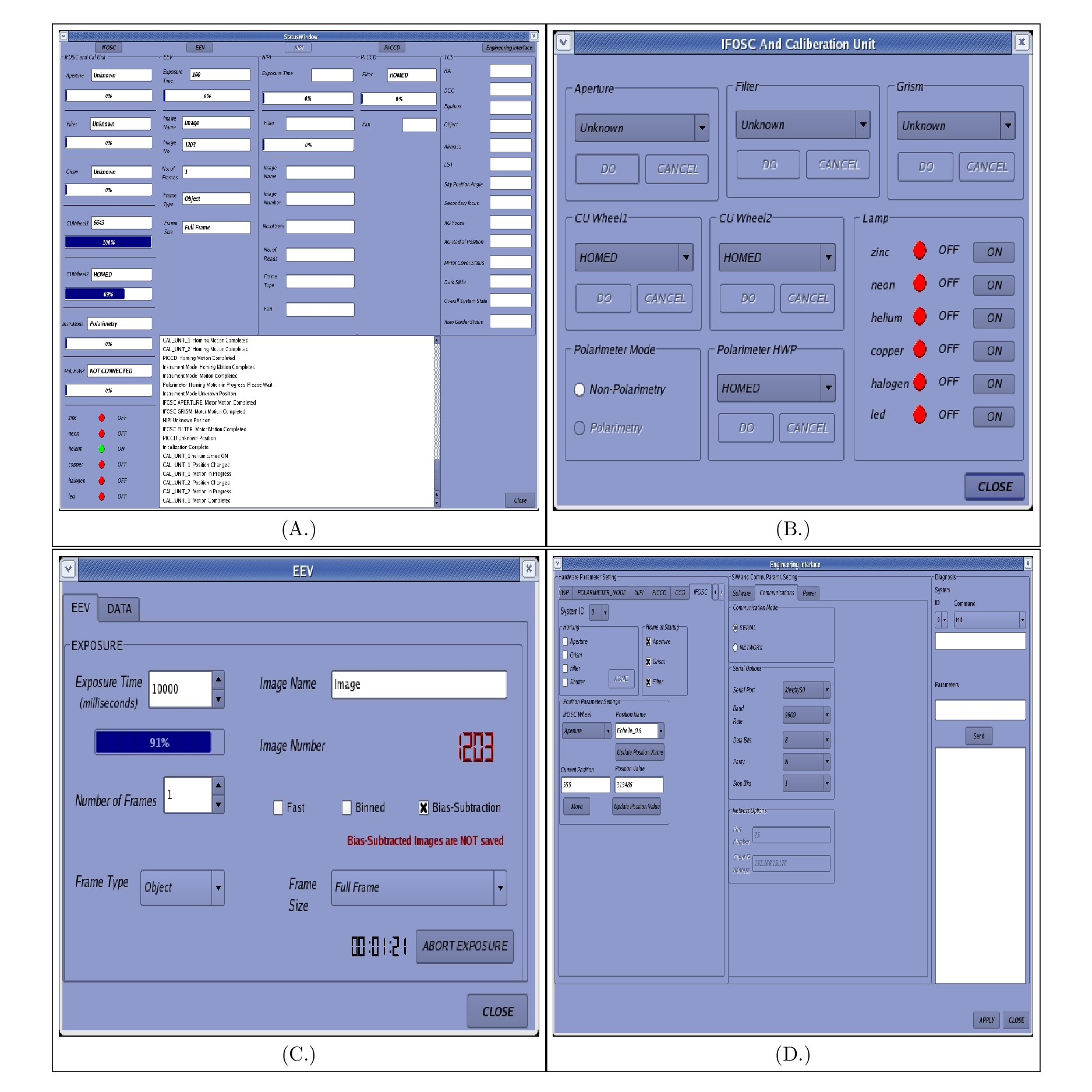}
\caption{The Graphical User's Interface of UNICS: (A.) Main window interface of GUI, showing the status of various subsystems (B.) Control interface of IFOSC (C.) Control and Status window for the data acquisition (D.) Engineering Interface of UNICS}
\label{UNICS-GUI}
\end{figure}

\begin{figure}
\centering
\epsscale{0.85}
\plotone{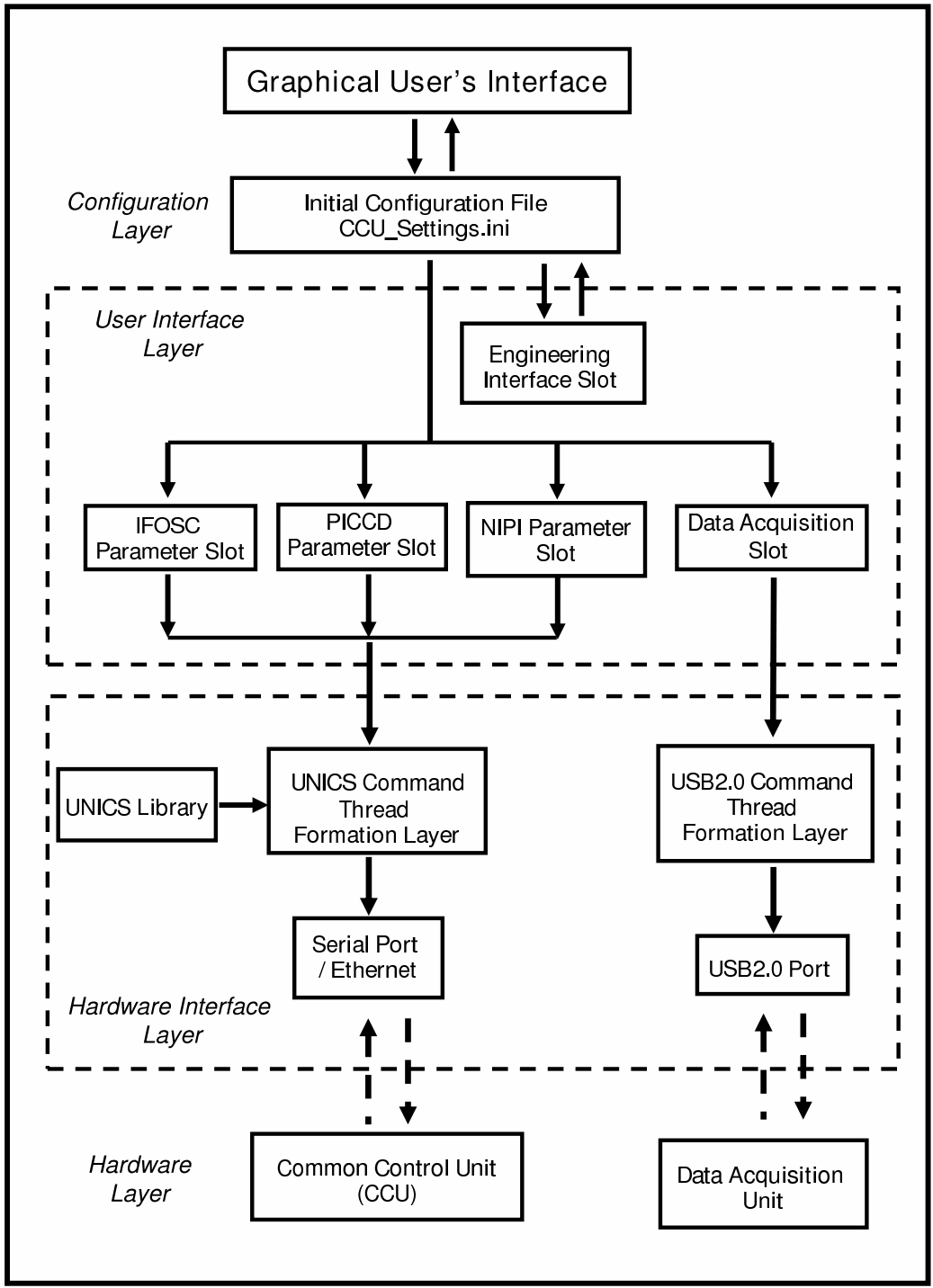}
\caption{Software Architecture of UNICS Application Software and GUI: The GUI combines the software controls of UNICS application programs and USB2.0 based data acquisition system.
}
\label{UNICS-Soft}
\end{figure}

At the front end of the UNICS, a Graphical User's Interface(GUI) is provided to take user's instructions and provide feedbacks regarding the state of the system and/or actions being currently executed (Figure~\ref{UNICS-GUI}). The GUI is developed in QT\footnote{http://trolltech.com/products/qt} and provides the real time visualization of various functions and system status. The status of various processes can be monitored on the GUI screen as they progress. It provides separate interface windows for different instruments and subsystems mounted on the Telescope. The control interface for IFOSC, PICCD, NIPI, and CCD data acquisition system are given through the main window of GUI. While working with any of these systems all the required parameters can be set through this interface and the tasks can be executed. An interface to SAO-DS9 image visualization package\footnote{http://hea-www.harvard.edu/RD/ds9} is given to GUI through XPA messaging system\footnote{http://hea-www.harvard.edu/RD/xpa/index.html} for the inspection of acquired images. The option of different image acquisition mode (e.g. Fast, Binned, and Bias Subtracted Mode) in the GUI helps the user to select appropriate mode while observing and thus minimizing the overhead time. However in general the GUI combines the controls of the following two separate hardware system in one graphical window (Figure~\ref{UNICS-Soft}): (1.) The UNICS Application software and (2.) The Application Software for the USB2.0 based data acquisition system.
\par
The UNICS Application Software converts the user's instructions to UNICS command structure and send to CCU through serial link or Ethernet network. The application software has been developed in {\it C++/C}. It supports multithreaded execution of tasks. This gives the flexibility to control a number of independent processes simultaneously. All the operation of UNICS Application Software (i.e. interaction of the GUI with CCU and data acquisition system) are controlled through entries in a configuration file named {\it $CCU\_settings.ini$}.  This configuration file is stored in ASCII text format and contains all the information describing each subsystem of the UNICS. The inclusion of any new subsystem to UNICS would then require addition of corresponding entries in this configuration file, and a corresponding interface window in GUI along with the underlying software of the subsystem. Also this file contains all the input parameters for serial communication protocol and Ethernet protocol. During the run time the entries in the configuration file keep on updating with the new values corresponding to change in any of the parameters of the subsystems. A library which consists of UNICS commands has also been developed. Any task executed via GUI would create a proper command packets using the data of {\it $CCU\_settings.ini$} and the UNICS library. These command packets would then be sent to CCU through serial link. The inbuilt Linux libraries are used for serial communication purpose.
\par
Apart from CCU, the GUI also provide access to the application software of USB2.0 based data acquisition system. It is an FPGA based system and communicate with the Server through USB2.0 interface. The same configuration file {\it $CCU\_settings.ini$} is used to interact with data acquisition system as is being used for the CCU. However the data acquisition system has a different library of its own and uses USB2.0 communication link for data transfer. The GUI also gives the interface between Telescope Control System (TCS) and the application software. During observations various observation parameters from the TCS can be read and displayed to the screen and can be included in the header of the output images. The Engineering Interface section of the GUI has been developed for the engineering level checks of the subsystems. It has the option to configure parameters of all basics commands of UNICS and transferring each of them to CCU in any desired sequence. Also this can be used to set the input parameters for Serial Communication Protocol and Ethernet Protocol. This feature has been proved to be of great utility while testing and debugging the system and access to engineering GUI is restricted through authentication due to safety concerns.

\section{Conclusion}

The UNified Instrument Control System (UNICS) has been developed keeping the general requirements of smaller and medium size observatories in picture. It provides a cost effective solution with simple hardware and software architecture to control and monitor the various instruments mounted on the Telescope. The hardware of UNICS is based on the ATmega 128 micro-controller and is designed on Master-Slave architecture while the Software architecture is based on the Linux Operating System. A user friendly GUI has been developed for the software control of UNICS which provides a good visibility of the current status of the system. The engineering mode of the GUI has been proved very useful in debugging hardware problems. Further considering the need for remote accessibility of UNICS, an option of replacing the serial communication link between the server and embedded Common Control Unit (CCU) on the Telescope by an Ethernet link has also been given to UNICS. The UNICS hardware offers good capability for expansion in future due to its generic architecture and interface approach. It greatly simplifies the issue of the control of current instruments and provides an easy solution to add more instrument and devices to the Telescope. Overall, UNICS is now a mature instrument control system that can handle existing and planned instruments and has the potential for future extensions. UNICS has been used on IUCAA 2m Telescope since March 2008.

\acknowledgements
We would like to thank Mr. Sunu Engineer, Mr. Ashwin Kumar, Ms. Ashwini Kadam, Mr. Moin Shaikh, and Mr. Sujit Punnadi for their support in the development of UNICS software. We are also thankful to other members of the Instrumentation Laboratory at IUCAA for their help and support during the development of UNICS. MKS thanks Council for Scientific and Industrial Research (CSIR), India, for the research grant award NO.9/545(25)/2005-EMR-I.

%%%%%%%%%%%%%%%%%%%%%%%%%%%%%%%%%%%%%%%%%%%%%%%%%%%%%%%%%
%%%%%%%%%%%%%%%%%%%%%%%%%%%%%%%%%%%%%%%%%%%%%%%%%%%%%%%%%

%%%%%%%%%%%%%%%%%%%%%%%%%%%%%%%%%%%%%%%%%%%%%%%%%%%%%%%%%
%%%%%%%%%%%%%%%%%%%%%%%%%%%%%%%%%%%%%%%%%%%%%%%%%%%%%%%%%
\clearpage

\clearpage

\clearpage

\clearpage

\clearpage

\end{document}